\documentstyle[twocolumn,aps]{revtex}

\begin{document}

\title{Semiclassical dynamical localization
and the multiplicative semiclassical propagator}
\author{L. Kaplan \thanks{kaplan@physics.harvard.edu}
\\Department of Physics and Society of Fellows,\\ Harvard
University, Cambridge, Massachusetts 02138}
\maketitle

\begin{abstract}
We describe an iterative approach to computing long-time semiclassical
dynamics in the presence of chaos,
which eliminates the need for summing over an exponentially
large number of classical paths, and has good convergence properties 
even beyond the Heisenberg time. Long-time semiclassical properties
can be compared with those of the full quantum system. The method is
used to demonstrate
semiclassical dynamical localization in one-dimensional classically
diffusive systems, showing that interference between classical paths
is a sufficient mechanism for limiting long-time phase space exploration.
\end{abstract}

\vskip 0.2in

Dynamical localization, the suppression of quantum phase space exploration in
a system whose classical analogue is diffusive, is a remarkable example
of non-ergodic behavior in quantizations of classically ergodic motion.
Since its discovery almost two decades ago,
the phenomenon has been discussed and observed in various numerical
studies\cite{ccif,numevidence},
and also in experimental settings\cite{exper}. Formal
connections with Anderson localization in disordered systems have been
made\cite{anders}.

As expected from classical-quantum correspondence,
diffusive quantum behavior is observed for times beyond the Ehrenfest time
in systems with
a diffusive classical limit. Localization then sets in (for dimensions
$d<2$) at a time scale
whose dependence on the diffusion constant $D$ and Planck's constant
$\hbar$ can be understood by analyzing when interference between 
classically distinct paths begins to be statistically important. It is
thus natural to ask whether phase interference between long classical
paths alone is sufficient to produce localization, in the
absence of ``hard quantum" effects like diffraction and tunneling. Addressing
this question has historically been made difficult by the exponential
proliferation of paths with time in a chaotic system. For $d=1$, for example,
the localization time scales as $D/\hbar^2$, so exact semiclassical calculations
all the way to the localization time scale
are in practice impossible to carry out for small values of $\hbar$, where
the semiclassical approximation itself is likely to be valid.

An attempt
along these lines in Ref.\cite{shudoikeda} proved somewhat inconclusive,
although some preliminary evidence of anomalous long-time behavior was found
at $21$--$27$ time steps. This in itself was an impressive calculational feat,
made possible by a symbolic dynamics and the piecewise-quadratic nature of the
potential. In Ref.~\cite{ss76,cohen}, statistical properties of
long periodic orbits were used to give plausibility arguments for
semiclassical localization without performing explicit periodic orbit
sums. In Ref.~\cite{ss77}, the relationship was examined
between the exact quantum propagator in a classically diffusive system,
and the semiclassical {\it one-step} propagator. It is important to
note, however, that iteration of a one-step semiclassical propagator
does not bear much resemblance to long-time semiclassical dynamics, except
insofar as both are (at least at short times) related to the quantum
dynamics\cite{multsc}.

In this work we adopt a different approach, based on the idea,
recently developed more fully in Ref.~\cite{multsc}, that even though
semiclassical propagation is not strictly multiplicative, long-time
semiclassical dynamics can in fact be well approximated by iteration of
intermediate-time propagators, with controllable errors and well-defined
convergence properties. In Ref.~\cite{multsc} this approach was used to
compute semiclassical dynamics past the Heisenberg time $T_H$ (where individual
eigenstates and eigenvalues can be resolved), for a system with $T_H=256$,
without exponential expenditure of computational effort.
Good convergence properties with decreasing $\hbar$ allowed
for direct comparison between quantum and semiclassical stationary
properties, such as long-time transport, spectra, and eigenstates.
In the present work we do not directly use any of the results of
Ref.~\cite{multsc}, but the interested reader is directed there for
a more complete discussion of the underlying ideas.

We begin by defining $A_t(i,j)$ to be the semiclassical propagator matrix 
taking quantum state $j$ to quantum state $i$ in time $t$ (computed
using the Gutzwiller--van Vleck semiclassical expression). Unlike
the corresponding quantum propagator $U_t(i,j)$, $A_t$ is not unitary, nor
is it multiplicative, e.g. $A_t \ne (A_{t/2})^2$. We can, however, easily
estimate the deviation from exact multiplicativity of the semiclassical
propagator, at least in the caustic-free case. We define the natural
basis-independent $L_2$ norm,
\begin{equation}
\label{norm}
\|A-B\|^2={1\over N} \sum_{i,j=1}^N |A_{ij}-B_{ij}|^2 \,,
\end{equation}
where the normalization ensures that the norm of a unitary operator is one.
We can then write
\begin{equation}
\label{oneerror}
\|A_t-(A_{t/2})^2\|^2 = O(\hbar^\alpha) \,,
\end{equation}
where for smooth dynamics, we have the exponent
$\alpha_{\rm smooth}=2$. This can be seen
by noting that $A_t$ is exactly given by combining two $A_{t/2}$ propagators,
as long as the intermediate integration at time $t/2$ is performed by
stationary phase. The relative error between performing the intermediate
integral exactly [$(A_{t/2})^2(x,y)=\int dz A_{t/2}(x,z)A_{t/2}(z,y)$]
and by stationary phase [$A_t(x,y)=\int_{\rm sp}dz A_{t/2}(x,z)A_{t/2}(z,y)$]
scales
as $\hbar$ (this being the order of the subleading term in the stationary
phase expansion); thus $\alpha_{\rm smooth}=2$\cite{multsc}.

In the case of a discontinuity in the underlying classical dynamics, which
is to be considered in the present work, the situation is rather
different. However, the general approach applies also to this case.
In expanding around each of the stationary paths that contribute
to $A_t(x,y)$, the region around the stationary phase point 
where the phase is slowly varying scales as
$\hbar^{1/2}$. Thus, for a stationary intermediate point $y$ within
$O(\hbar^{1/2})$ of a discontinuity in the potential (or in its first
derivative), the exact integral gets cut off within the region of gaussian
integration, and the relative error between the full integral and
the stationary phase approximation
is of order unity. It is easy to see that for small
$\hbar$ this diffractive effect dominates
the effect of the subleading terms in the expansion (which as we saw lead to
$\alpha_{\rm smooth}=2$), and results in
\begin{equation}
\label{diffrerr}
\alpha_{\rm discontinuous}=1/2\,.
\end{equation}
For long times $t$, many classical paths must be summed over to obtain the
semiclassical propagator; 
here we are of course assuming that errors in the sum over paths
add no more coherently
than the actual contributions themselves.

The next step is to extend Eq.~\ref{oneerror} to the more general form
\begin{equation}
\label{manyerror}
\|A_t-(A_{t/M})^M\|^2 = O(M\hbar^\alpha) = O\left(\left({t \over M}\right)^{-1}
t \hbar^\alpha \right) \,,
\end{equation}
which follows from assuming the successive errors in replacing $M-1$
stationary phase integrals by exact ones to add incoherently. The assumption
of incoherent addition of errors
breaks down for very large $M$ (specifically, for $M$ greater than 
$\hbar^{-1}$, the
Heisenberg time measured in units of the shortest periodic orbit\cite{multsc}).
However,
the higher-order corrections in $M$ will not be relevant for our purposes.

Eq.~\ref{manyerror} allows successive controlled approximations to be
computed to the exact semiclassical dynamics by taking $t/M \gg 1$ (this
produces values much closer to the semiclassical than to the quantum
results). Of course,
taking $M \to \infty$ ($t/M \ll 1$) in the expression $(A_{t/M})^M$,
we instead recover the quantum
propagator, as in the Feynman path integral formalism. The intermediate case
$t/M \sim 1$ (as in the Bogomolny surface of section approach\cite{bogo})
produces a long-time dynamics which is strictly speaking neither quantum
nor semiclassical, and provides an interpolation between the two worlds.

The scaling properties of the iterative semiclassical
approximation with $\hbar$, time $t$, 
and ``quantization time" $T_Q \equiv t/M$, as expressed in Eq.~\ref{manyerror}
above, hold even for times $t$ beyond the Heisenberg time of the system. These
scaling properties, based on power-counting arguments,
have been extensively tested numerically in \cite{multsc}. One qualification
is that while for $\alpha=2$ and $T_H \sim \hbar^{-1}$ the approximation 
at the Heisenberg time using fixed $T_Q$ gets better and better as
$\hbar \to 0$, in the present situation
we have $\alpha=1/2$ and $T_H \sim \hbar^{-2}$, so as $\hbar$ gets small
we need larger $T_Q$ to preserve the accuracy of the approximation.

We are now ready to apply the above outlined formalism to the case at hand:
dynamical localization in one-dimensional systems. We consider a kicked
map\cite{ccif,kickmap} on a cylindrical phase space $0 \le q < 2\pi$,
$-\infty < p < \infty$,
\begin{eqnarray}
\label{clasdyn}
\tilde p & = & p - V'(q) \nonumber \\
\tilde q & = & q + \tilde p \; {\rm mod} \; 2\pi \,,
\end{eqnarray}
with kick potential 
\begin{equation}
\label{kickpot}
V(q) = - {1 \over 2}K (q-\pi)^2+B \cos q \;\;\;(0 \le q < 2\pi)
\end{equation}
turned on momentarily once every time step. Locally the dynamics
looks everywhere like an inverted harmonic oscillator with a sinusoidal
perturbation (as long as $B<K$), except for a discontinuity in the impulse
at $q=0$.
The classical motion is completely chaotic, and diffusive in $p$,
\begin{equation}
\label{clasdiff}
\langle (p-p_0)^2 \rangle_{\rm classical}=Dt \,,
\end{equation}
with diffusion constant
\begin{equation}
\label{dval}
D=\langle (V')^2 \rangle =
{1 \over 2} \left[\left({2 \pi^2 K^2 \over 3}\right) 
+B^2-4KB\right] \,.
\end{equation}

The quantization of Eq.~\ref{clasdyn} is straightforward\cite{ccif,kickmap}.
Choosing periodic boundary conditions in $q$-space, we have a momentum
basis given by $p_n=n\hbar$, $n=-\infty \ldots \infty$. The dynamics is
given by a unitary one-step propagator
\begin{equation}
\label{quandyn}
U=e^{-i\hat p^2 / 2\hbar}e^{-iV(\hat q) / \hbar} \,.
\end{equation}

Because the quantum dynamics (as well as the classical) is symmetric under
parity [$p \to -p$, $q \to 2\pi -q$], we will in what follows focus only on the even sector $|p\rangle_{\rm even}=(|p\rangle+|-p\rangle)/\sqrt 2$, $p>0$.
This eliminates the problem of tunneling between positive and negative
momenta.                 

The semiclassical dynamics (in the absence of caustics and Maslov phases,
which are conveniently avoided by taking $B<K$ in Eq.~\ref{kickpot})
is given by the standard Gutzwiller-van Vleck propagator
\begin{eqnarray}
\label{gvv}
A_{\rm sc}(p',p,t) & = & \left[{1 \over 2\pi i\hbar}\right]^{d/2}
\sum_j \left|\det {\partial^2 S_j(p,p',t) \over \partial p \partial p'}
\right|^{1/2} \nonumber \\ & \times &
\exp{i S_j(p,p',t) \over \hbar}
\,,
\end{eqnarray}
where $S_j$ is the action for classical path $j$ taking $p$ to $p'$,
and the determinant
is the corresponding classical probability density.
We can now use Eq.~\ref{gvv} to evaluate the semiclassical propagator matrices
$A_{T_Q}$ for various ``quantization times'' $T_Q$ \cite{multsc}, and then
iterate to obtain $A_t \approx (A_{T_Q})^{t/T_Q}$. As $T_Q \to t$,
we obtain the exact semiclassical behavior. In general, though,
we only need to take $T_Q$ large enough to obtain the desired level of
convergence to the true long-time behavior. The behavior of the
iterative semiclassical limit as $T_Q$ becomes large can be compared with
the quantum dynamics as given by $U_t=U^t$ (Eq.~\ref{quandyn}).

The numerical results of this localization
study are presented in Figs.~\ref{fig1},~\ref{fig2}.
In Fig.~\ref{fig1}, we choose a piecewise linear map, with parameters $K=1.073$,
$B=0.0$ in the kick potential of Eq.~\ref{kickpot}.
In Fig.~\ref{fig2}, a sinusoidal term is added to
the potential: $K=1.073$, $B=0.52$. In both cases, the quantum and
semiclassical calculations are performed with $\hbar=0.293$ (note that this
takes us well into the semiclassical regime: the relevant expansion parameter
is $\hbar/(2\pi)^2$, since $(2\pi)^2$ is the area of a unit cell in phase
space). We can
now compute $\langle (n-n_0)^2\rangle$, the spread in momentum space
in units of $\hbar$, as a function of time ($n=p/\hbar$).

More explicitly, given a propagator $G_t$, whether quantum or semiclassical,
we define
\begin{equation}
\label{pspread}
\langle (n-n_0)^2\rangle = \left\langle{\sum_i |\langle i|G_t|j\rangle|^2
\left ( {p_i \over \hbar} - {p_j \over \hbar} \right )^2 \over
\sum_i |\langle i|G_t|j\rangle|^2}\right\rangle_j
\end{equation}
The average over $j$
is performed over initial momenta far from $0$ and also far from the edge
of the numerical lattice.

The classical
diffusion result given by Eqs.~\ref{clasdiff},~\ref{dval} appears on the
log-log plot in Figs.~\ref{fig1},~\ref{fig2}
as a straight line of slope one. The full quantum calculation
is seen in the dashed curve, which in each case is seen to turn over and
approach a constant after the localization time $T_{\rm loc}=D/\hbar^2$.
This theoretically expected value of $T_{\rm loc}$ (which is also
equal to the predicted RMS
spread in momentum at infinite time in units of $\hbar$, the square root 
of the quantity plotted in Figs.~\ref{fig1},\ref{fig2}) is given by
$T_{\rm loc}=44.1$ in Fig.~\ref{fig1} and $T_{\rm loc}=32.7$ in
Fig.~\ref{fig2}. All these predictions are in reasonable
agreement with the full quantum
numerics.

We now proceed to the semiclassical analysis. The dotted curve in each of the
two Figures represents the momentum
spreading given by iteration of the one-step
semiclassical propagator $A_1$. This does not closely follow the full
quantum (or, as we shall soon see, the exact semiclassical result), and
the behavior of this quantity in Fig.~\ref{fig2} is particularly erratic.

Now,
guided by Eq.~\ref{manyerror}, we look for convergence to the
exact semiclassical answer as the quantization time $T_Q=t/M$ is taken to be
much greater than one. Specifically, in Fig.~\ref{fig1} we plot as solid lines
the calculations with $T_Q=7,8,9$ (i.e. we take successive approximations
$A_t \approx (A_7)^{t/7}$, etc.). We see that the agreement between the
three calculations is very good, strongly suggesting that convergence
has been achieved. The semiclassical calculation begins to deviate
from the quantum sometime around $T_{\rm loc}$; nevertheless it does
very clearly localize at a well defined momentum spread somewhat larger
than that given by the quantum calculation. (Some difference in the
details of the end of classical diffusion for the quantum and
semiclassical calculations is not surprising.
One should note here for example that
the discontinuity in the kick potential will have a diffractive effect on the
quantum dynamics, one which will not be present in the semiclassical
approximation.)

In Fig.~\ref{fig2}, the $T_Q \to \infty$ convergence to the exact
long-time semiclassical dynamics is found to be somewhat
slower (calculations with $T_Q=8,9,10$ are plotted as solid lines).
Nevertheless, up until $t \approx 200 \approx 6T_{\rm loc}$, the three
curves are in very 
good agreement, with uncertainty small not only compared to their
common deviation from classical diffusion (straight line), but also
compared to their
common distance from the quantum curve (dashed).
The evidence for localization is
very clear in this case also,
as is the failure of the one-step iterative approximation
(dotted curve) to reproduce long-time semiclassical behavior.

As an additional test of semiclassical localization at very long times,
we consider the eigenstates of the successive propagators $A_{T_Q}$ as
$T_Q \to \infty$. In the absence of interference effects (i.e. considering
each $A_{T_Q}$ simply as a band random matrix of band width
$\sqrt{DT_Q}/\hbar$), we would expect the momentum spread $(\delta p)^2$
of the typical eigenstate to increase linearly with $T_Q$. In fact, however,
phase interference between classical paths turns out to be very important
indeed, and the average RMS width of the eigenstates of $A_{T_Q}$ is found
not to increase significantly with $T_Q$ once $T_Q \gg 1$. In Fig.~\ref{fig3},
the mean RMS width (in units of $\hbar$) of $A_{T_Q}$ eigenstates centered well
away from the edge of the numerical lattice is plotted vs. the quantization
time $T_Q$. Parameters are the same as in the previous two figures: 
plusses are used for the $B=0.0$ case (corresponding to Fig.~\ref{fig1})
and squares for $B=0.52$ (as in Fig.~\ref{fig2}). The result using
the $T_Q=1$ one-step iterated semiclassical propagator is also plotted,
as is the quantum momentum spread at $T_Q=0$. We see localization at
large $T_Q$ for both sets of parameters; in each case, the localization length
is somewhat larger semiclassically as compared to the quantum calculation.
The result obtained using $T_Q=1$ is intermediate between semiclassical
and quantum in both cases.

We can test the scaling of the semiclassical localization length with the
diffusion constant by comparing the results for $B=0.0$ and $B=0.52$.
The ratio of semiclassical localization lengths for these two parameters
is found to be $1.32$ (using $T_Q=6$); analytically we predict $1.35$.

The behavior of the semiclassical localization with $\hbar$ has also been
investigated, and we find that the localization length increases
roughly in accordance with the theoretical predictions of dynamical
localization theory. However, the semiclassical localization length
does grow somewhat more slowly with $1/\hbar$ than the quantum localization
length, apparently leading to a convergence between these quantities at small
$\hbar$. Thus, for $B=0.52$ and using $T_Q=6$ we find a ratio of
$1.66$ between the long-time semiclassical and quantum localization lengths;
reducing $\hbar$ by a factor of $2$ causes this ratio to drop to $1.35$.
Unfortunately we were not able to investigate extremely small $\hbar$ due to
computer limitations.

Thus, using an iterative approach to long-time semiclassical
calculations, we have been able to see explicitly semiclassical
localization in classically diffusive systems at small $\hbar$.
We can now say
definitively that although details of long-time quantum dynamics are
affected by diffraction and tunneling corrections, the essence of the
localization phenomenon is indeed contained in the interference among
long classical paths.

This research was supported by the National Science Foundation under
Grant No. 66-701-7557-2-30. Some of the work was performed during a stay at
the Weizmann Institute in Israel. The author is very grateful to E. J. Heller
for many useful discussions on semiclassical methods.

\begin{figure}
\caption{
Momentum spread $\langle (n-n_0)^2 \rangle$ (as defined by Eq.~\ref{pspread}),
as a function of time for a kicked system with kick potential parameters
$K=1.073$, $B=0.0$ (Eq.~\ref{kickpot}). Classical diffusion
(Eq.~\ref{clasdiff})
appears as a straight line; the quantum calculation,
represented by a dashed curve, shows dynamical localization at time
scale $T_{\rm loc}=44.1$. Successive approximations to the long-time
semiclassical propagator (using $T_Q=7,8,9$) are drawn as solid curves.
The one-step iterated semiclassical propagator ($T_Q=1$) produces
the dotted curve.
}
\label{fig1}
\end{figure}

\begin{figure}
\caption{
Same as previous Figure, with sinusoidal perturbation in the potential,
$B=0.52$. Here the expected value of $T_{\rm loc}$ is $32.7$.
Approximations to long-time semiclassical propagation using $T_Q=8,9,10$
are shown as solid curves.
}
\label{fig2}
\end{figure}

\begin{figure}
\caption{
Average RMS eigenstate width for successive approximations to long-time
semiclassical propagation, $T_Q=3 \ldots 9$. The average eigenstate
width for the one-step iterated propagator ($T_Q=1$) is also displayed,
as is the quantum result ($T_Q=0$). Plusses represent the case $B=0.0$;
squares represent $B=0.52$.
}
\label{fig3}
\end{figure}

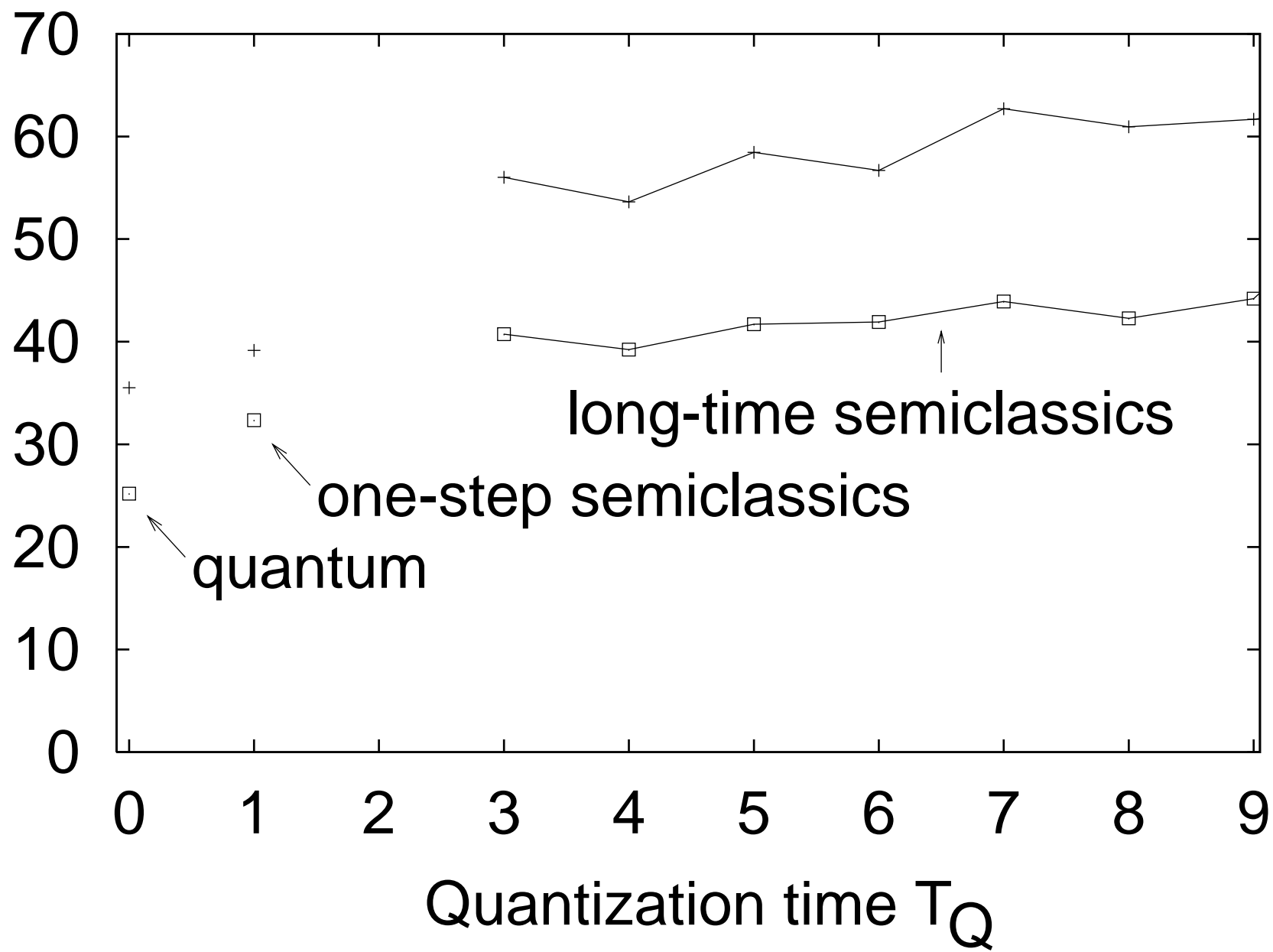
\end{document}